\documentstyle[aps,preprint]{revtex}

\begin{document}
\draft
\title{Kawasaki-type Dynamics: Diffusion in the kinetic Gaussian model}
\author{Han Zhu$^2$ and Jian-Yang Zhu$^{1,3\thanks{%
Author to whom correspondence should be addressed. Address correspondence to
Department of Physics, Beijing Normal University, Beijing 100875, China.
Email address: zhujy@bnu.edu.cn}}$}
\address{$^1$CCAST (World Laboratory), Box 8730, Beijing 100080, China\\
$^2$Department of Physics, Nanjing University, Nanjing, 210093, China\\
$^3$Department of Physics, Beijing Normal University, Beijing 100875, China}
\maketitle

\begin{abstract}
In this article, we retain the basic idea and at the same time generalize
Kawasaki's dynamics, spin-pair exchange mechanism, to spin-pair
redistribution mechanism, and present a normalized redistribution
probability. This serves to unite various order-parameter-conserved
processes in microscopic, place them under the control of a universal
mechanism and provide the basis for further treatment. As an example of the
applications, we treated the kinetic Gaussian model and obtained exact
diffusion equation. We observed critical slowing down near the critical
point and found that, the critical dynamic exponent $z=1/v=2$ is independent
of space dimensionality and the assumed mechanism, whether Glauber-type or
Kawasaki-type.
\end{abstract}

\pacs{PACS number(s): 64.60.Ht, 05.70.Ln, 75.10.Hk, 68.35.Fx}

Irreversible dynamic systems exhibit complicated but interesting
nonequilibrium phenomena near the critical point. In spite of their
complexity, the interesting dynamic critical behaviors have been attracting
a lot of researchers for many decades. Within the vast body of literature,
two pioneering works completed by Glauber\cite{Glauber} and Kawasaki\cite
{Kawasaki} have been regarded as a milestone. Great progress has been
achieved with the application of Glauber's single-spin flip mechanism and
kawasaki's spin-pair exchange mechanism. These two schemes catch the
inherent essential process bound in.

Kawasaki's dynamics deals with a system consisting of an array of $N$
coupled spins. The coupling between spins is represented by a set of
transition probabilities of spin exchange. Initially, the focus was on Ising
model. The spins may exchange with their nearest neighbors, and in this way,
the system evolves while the total spin remains conserved. In later studies
the idea of exchange has been proved very successful as it catches the
essential nature. It has important applications in Ising model\cite
{Ising-1,Ising-2} and Ising-like models such as lattice gas model\cite
{Lat-Gas-1,Lat-Gas-2,Lat-Gas-3}, Blume-Emery-Griffiths (BEG) model\cite
{BEG-1,BEG-2} and others\cite{Others-1,Others-2}. These applications have
contributed a lot to the understanding of the thermodynamics, and have
erected itself as the basic mechanism in the class of
order-parameter-conserved processes.

However, its initial embodiments was closely tied up with the simpleness of
Ising model, in which the spins can only take two values, $\pm 1$, and there
exists only nearest-neighbor interactions. It turned to be limited when
applied to systems other than Ising or Ising-like models. The same situation
exists in Glauber's dynamics. Recently Zhu and Yang successfully generalized
Glauber's single-spin flip mechanism to single-spin transition mechanism and
gave a normalized version of the transition probability\cite{Zhu-1}. The
applications in several models have yielded encouraging results\cite
{Zhu-1,Zhu-2}. In this article, along the same line, we retain the basic
idea and at the same time generalize Kawasaki's dynamics, spin-pair exchange
mechanism, to spin-pair redistribution mechanism, and also present a
normalized redistribution probability.

The limit is to be removed and the spins may take various discrete values
(discrete-spin model) or continuous values (continuous-spin model). Because
in Ising model spins can only have two opposite values, simply direct
exchange may be enough to describe the way the system evolves. In other
different models, such as Gaussian model, Potts model, or XY model, this
simple picture may not be as capable. On the other hand, as is mentioned
above, the conservation of the order-parameter has been regarded as the most
important feature of this class of processes (and a necessary result of the
exchange mechanism). Based on these considerations, in the generalized
mechanism, spin pair redistribution mechanism, two neighboring spins, $%
\sigma _k$, $\sigma _{k+1}$, no longer merely exchange with each other.
Instead, they may take any values, $\hat{\sigma}_k$, $\hat{\sigma}_{k+1}$,
as long as their sum remains conserved, or, their sum is redistributed; and
we call this, spin-pair redistribution.

The probability distribution function $P\left( \sigma _1,\ldots ,\sigma
_N;t\right) $, or simply $P\left( \left\{ \sigma \right\} ;t\right) $,
denotes the probability of the $N$-spin system being in the state $\left(
\sigma _1,\ldots ,\sigma _N\right) $, or simply $\left\{ \sigma \right\} $,
at time $t$. The $j$th and $l$th spins are two neighboring spins, and let $%
W_{jl}\left( \sigma _j\sigma _l\rightarrow \hat{\sigma}_j\hat{\sigma}%
_l\right) $ be the probability per unit time that they are redistributed
while the others remain unchanged. Then, on the supposition of neighboring
spin-pair redistribution, we may write the time derivative of the function $%
P\left( \left\{ \sigma \right\} ;t\right) $ as 
\begin{eqnarray}
\frac d{dt}P\left( \left\{ \sigma \right\} ;t\right) &=&\sum_{\left\langle
jl\right\rangle }\sum_{\hat{\sigma}_j,\hat{\sigma}_l}\left\{ -W_{jl}\left(
\sigma _j\sigma _l\rightarrow \hat{\sigma}_j\hat{\sigma}_l\right) P\left(
\left\{ \sigma \right\} ;t\right) \right.  \nonumber \\
&&\left. +W_{jl}\left( \hat{\sigma}_j\hat{\sigma}_l\rightarrow \sigma
_j\sigma _l\right) P\left( \left\{ \sigma _{i\neq j,l}\right\} ,\hat{\sigma}%
_j,\hat{\sigma}_l;t\right) \right\}  \label{Master eq}
\end{eqnarray}
This is a probability equation, in which the first term in the right-hand
side of (\ref{Master eq}) denotes the decrease of the probability
distribution function $P\left( \left\{ \sigma \right\} ;t\right) $ per unit
time due to the redistribution of the spin pair from initially $\sigma
_j\sigma _l$ to various values $\hat{\sigma}_j\hat{\sigma}_l$, and the
second term denotes the increase of $P\left( \left\{ \sigma \right\}
;t\right) $ due to the redistribution of the spin pair from initially
various values $\hat{\sigma}_j\hat{\sigma}_l$ to finally $\sigma _j\sigma _l$%
. (Clearly, in Ising and Ising-like systems, it does become Kawasaki's
picture.) We shall refer to (\ref{Master eq}) as the master equation since
its solution would contain the most complete description of the system
available.

The key step for solving the master equation is the determination of the
redistribution (or exchange, flip, transition, etc.) probability. Usually it
can not be uniquely determined by the detailed balance condition, and thus,
in Kawasaki's as well as Glauber's pioneering works, some arbitrariness
remained. We hope to make our choice of the spin-pair redistribution
probability able to contain Kawasaki's in the specific Ising model, and
applicable to various other spin systems, while at the same time
appropriate, clearer, and more definite. Now we consider it in both
mathematical and physical aspects. In mathematics, generally speaking, the
probability must be positive and able to be normalized; in physics, we often
require that a system in thermodynamic equilibrium satisfy the detailed
balance condition; we also require the probability be ergodic as long as the
total spin keeps conserved through the redistribution. Based on these
considerations, for arbitrary neighboring $j$th and $l$th spins, we can
choose the redistribution probability $W_{jl}\left( \sigma _j\sigma
_l\rightarrow \hat{\sigma}_j\hat{\sigma}_l\right) $ to satisfy the following
conditions:

(a) Ergodicity, positivity, and conservation of spin: 
\[
W_{jl}\left( \sigma _j\sigma _l\rightarrow \hat{\sigma}_j\hat{\sigma}%
_l\right) >0,~\forall \hat{\sigma}_j+\hat{\sigma}_l=\sigma _j+\sigma _l 
\]
\[
W_{jl}\left( \sigma _j\sigma _l\rightarrow \hat{\sigma}_j\hat{\sigma}%
_l\right) =0,~\forall \hat{\sigma}_j+\hat{\sigma}_l\neq \sigma _j+\sigma _l 
\]
this may suggest a $\delta $ function in the expression;

(b) Normalization: 
\[
\sum_{\hat{\sigma}_j,\hat{\sigma}_l}W_{jl}\left( \sigma _j\sigma
_l\rightarrow \hat{\sigma}_j\hat{\sigma}_l\right) =1; 
\]

(c) Detailed balance: 
\[
\frac{W_{jl}\left( \sigma _j\sigma _l\rightarrow \hat{\sigma}_j\hat{\sigma}%
_l\right) }{W_{jl}\left( \hat{\sigma}_j\hat{\sigma}_l\rightarrow \sigma
_j\sigma _l\right) }=\frac{P_{eq}\left( \sigma _1,\ldots ,\hat{\sigma}_j,%
\hat{\sigma}_l,\ldots ,\sigma _N\right) }{P_{eq}\left( \sigma _1,\ldots
,\sigma _j,\sigma _l,\ldots ,\sigma _N\right) }, 
\]
in which 
\[
P_{eq}=\frac 1Z\exp \left[ -\beta {\cal H}\left( \left\{ \sigma \right\}
\right) \right] ,~Z=\sum_{\left\{ \sigma \right\} }\exp \left[ -\beta {\cal H%
}\left( \left\{ \sigma \right\} \right) \right] , 
\]
where $P_{eq}$ is the equilibrium Boltzmann distribution function, $Z$ is
the partition function, and ${\cal H}\left( \left\{ \sigma \right\} \right) $
is the system Hamiltonian.

Although the spin pair redistribution probabilities still are not determined
uniquely by the above restriction, there is less room left. The
consideration that we have is similar to that in the generalization of
Glauber's dynamics\cite{Zhu-1}, that the redistribution of a neighboring
pair depends merely on the momentary values of the spins surrounding them,
or, the spins of the system, and the influence of the heat bath. Based on
this, we can similarly assume that the redistribution probability $%
W_{jl}\left( \sigma _j\sigma _l\rightarrow \hat{\sigma}_j\hat{\sigma}%
_l\right) $ only depends on the heat Boltzmann factor of the system. 
\[
W_{jl}\left( \sigma _j\sigma _l\rightarrow \hat{\sigma}_j\hat{\sigma}%
_l\right) \propto \delta _{\sigma _j,\sigma _l;\hat{\sigma}_j,\hat{\sigma}%
_l}\exp \left[ -\beta {\cal H}_{jl}\left( \hat{\sigma}_j,\hat{\sigma}%
_l,\left\{ \sigma _m\right\} _{m\neq j,l}\right) \right] 
\]
or 
\begin{equation}
W_{jl}\left( \sigma _j\sigma _l\rightarrow \hat{\sigma}_j\hat{\sigma}%
_l\right) =\frac 1{Q_{jl}}\delta _{\sigma _j,\sigma _l;\hat{\sigma}_j,\hat{%
\sigma}_l}\exp \left[ -\beta {\cal H}_{jl}\left( \hat{\sigma}_j,\hat{\sigma}%
_l,\left\{ \sigma _m\right\} _{m\neq j,l}\right) \right] .  \label{red-pro}
\end{equation}
Using the normalization condition (b), we can write the normalization factor 
$Q_{jl}$ as 
\[
Q_{jl}=\sum_{\hat{\sigma}_j,\hat{\sigma}_l}\delta _{\sigma _j,\sigma _l;\hat{%
\sigma}_j,\hat{\sigma}_l}\exp \left[ -\beta {\cal H}_{jl}\left( \hat{\sigma}%
_j,\hat{\sigma}_l,\left\{ \sigma _m\right\} _{m\neq j,l}\right) \right] . 
\]
In fact, the Hamiltonian coming from the interaction of spins unrelated to $%
\sigma _j,\sigma _l$ will be cancelled, and thus in the actual calculation,
one needs only write the effective Hamiltonian. Obviously, $Q_{jl}$ is
related to the temperature, the surrounding spins and the conserved sum of $%
\sigma _j$, $\sigma _l$.

Compared with Kawasaki's expression, (\ref{red-pro}) is a normalized
version. In his expression, there is an $\alpha $ appearing in the exchange
probability, which was assumed to be a constant. In Ising model,
redistribution is in fact exchange and actually our expression is only a
definite selection for constant $\alpha $ by extra restriction and physical
considerations.

Usually, we are interested in local magnetization. It is defined as 
\begin{equation}
q_k\left( t\right) =\left\langle \sigma _k\left( t\right) \right\rangle
=\sum_{\left\{ \sigma \right\} }\sigma _kP\left( \left\{ \sigma \right\}
;t\right) .  \label{q}
\end{equation}
According to the definition (\ref{q}) and the master equation (\ref{Master
eq}), and using the normalization condition (b), the time-evolving equation
of $q_k\left( t\right) $ can be derived as (see Appendix \ref{App-A}) 
\begin{equation}
\frac d{dt}q_k\left( t\right) =-2dq_k\left( t\right) +\sum_{\left\{ \sigma
\right\} }\sum_w\left[ \sum_{\hat{\sigma}_k,\hat{\sigma}_{k+w}}\hat{\sigma}%
_kW_{k,k+w}\left( \sigma _k\sigma _{k+w}\rightarrow \hat{\sigma}_k\hat{\sigma%
}_{k+w}\right) \right] P\left( \left\{ \sigma \right\} ;t\right) ,
\label{Time-Q}
\end{equation}
where $d$ is the dimensionality of the system, and $\sum_w$ means summation
over the nearest neighbors (clearly it is related to the dimensionality,
too).

Kawasaki's exchange mechanism was initially designed for the study of
diffusion constant, and he himself obtained an approximate result for Ising
model by first deriving an expression of spin flux\cite{Kawasaki}. Now we
will treat the kinetic Gaussian model as an application of our
redistribution mechanism, while our method is a direct one.

The Gaussian model, proposed by T. H. Berlin and M. Kac, at first in order
to make an Ising model more tractable, is an continuous-spin model. It has
the same Hamiltonian form as the Ising model (three dimensional), 
\begin{equation}
-\beta {\cal H}=K\sum_{i,j,k=1}^N\sum_w\sigma _{ijk}\left( \sigma
_{i+w,jk}+\sigma _{ij+w,k}+\sigma _{ij,k+w}\right) ,  \label{G-Ham}
\end{equation}
where $\sum_w$ means summation over nearest neighbors. Comparing it with the
Ising model, there are two extensions: First, the spins $\sigma _{ijk}$ can
take any real value between $\left( -\infty ,+\infty \right) $. Second, to
prevent the spins from tending to infinity, the probability of finding a
given spin between $\sigma _{ijk}$ and $\sigma _{ijk}+d\sigma _{ijk}$ is
assumed to be the Gaussian-type distribution 
\begin{equation}
f\left( \sigma _{ijk}\right) d\sigma _{ijk}=\sqrt{\frac b{2\pi }}\exp \left(
-\frac b2\sigma _{ijk}^2\right) d\sigma _{ijk},  \label{Spin-dis}
\end{equation}
where $b$ is a distribution constant independent of temperature. Although it
is an extension of the Ising model, the Gaussian model is quite different.
In the equilibrium case, on translational invariant lattices the Gaussian
model was exactly solvable, and later as a starting point to study the
unsolvable models it was also investigated with mean field theory and the
momentum-space renormalization-group method.

In the $3D$ kinetic Gaussian model, the system Hamiltonian and the spin
distribution probability are (\ref{G-Ham}) and (\ref{Spin-dis}),
respectively. In this case there are six combined terms in the $3D$ type
time-evolving equation of the local magnetization, Eqs.(\ref{Time-Q}), and
the details of these complex calculations are in Appendix \ref{App-B}. Here
we give only the results, 
\begin{eqnarray*}
&&\sum_{\hat{\sigma}_{ijk},\hat{\sigma}_{i+1,j,k}}\hat{\sigma}%
_{ijk}W_{i,j,k;i+1,j,k}\left( \sigma _{ijk}\sigma _{i+1,j,k}\rightarrow \hat{%
\sigma}_{ijk}\hat{\sigma}_{i+1,j,k}\right) \\
&=&\frac 1{2\left( b+K\right) }\left[ K\left( \sigma _{i,j+1,k}+\sigma
_{i-1,j,k}+\sigma _{i,j-1,k}+\sigma _{i,j,k-1}+\sigma _{i,j,k+1}+\sigma
_{ijk}+\sigma _{i+1,j,k}\right. \right. \\
&&\left. \left. -\sigma _{i+1,j+1,k}-\sigma _{i+2,j,k}-\sigma
_{i+1,j-1,k}-\sigma _{i+1,j,k+1}-\sigma _{i+1,j,k-1}\right) +b\left( \sigma
_{ijk}+\sigma _{i+1,j,k}\right) \right] ,
\end{eqnarray*}
\begin{eqnarray*}
&&\sum_{\hat{\sigma}_{ijk},\hat{\sigma}_{i-1,j,k}}\hat{\sigma}%
_{ijk}W_{i,j,k;i-1,j,k}\left( \sigma _{ijk}\sigma _{i-1,j,k}\rightarrow \hat{%
\sigma}_{ijk}\hat{\sigma}_{i-1,j,k}\right) \\
&=&\frac 1{2\left( b+K\right) }\left[ K\left( \sigma _{i,j+1,k}+\sigma
_{i+1,j,k}+\sigma _{i,j-1,k}+\sigma _{i,j,k-1}+\sigma _{i,j,k+1}+\sigma
_{ijk}+\sigma _{i-1,j,k}\right. \right. \\
&&\left. \left. -\sigma _{i-1,j+1,k}-\sigma _{i-2,j,k}-\sigma
_{i-1,j-1,k}-\sigma _{i-1,j,k+1}-\sigma _{i-1,j,k-1}\right) +b\left( \sigma
_{ijk}+\sigma _{i-1,j,k}\right) \right] ,
\end{eqnarray*}
\begin{eqnarray*}
&&\sum_{\hat{\sigma}_{ijk},\hat{\sigma}_{i,j+1,k}}\hat{\sigma}%
_{ijk}W_{i,j,k;i,j+1,k}\left( \sigma _{ijk}\sigma _{i,j+1,k}\rightarrow \hat{%
\sigma}_{ijk}\hat{\sigma}_{i,j+1,k}\right) \\
&=&\frac 1{2\left( b+K\right) }\left[ K\left( \sigma _{i,j,k+1}+\sigma
_{i,j-1,k}+\sigma _{i,j,k-1}+\sigma _{i-1,j,k}+\sigma _{i+1,j,k}+\sigma
_{ijk}+\sigma _{i,j+1,k}\right. \right. \\
&&\left. \left. -\sigma _{i,j+1,k+1}-\sigma _{i,j+2,k}-\sigma
_{i,j+1,k-1}-\sigma _{i+1,j+1,k}-\sigma _{i-1,j+1,k}\right) +b\left( \sigma
_{ijk}+\sigma _{i,j+1,k}\right) \right] ,
\end{eqnarray*}
\begin{eqnarray*}
&&\sum_{\hat{\sigma}_{ijk},\hat{\sigma}_{i,j-1,k}}\delta _{\hat{\sigma}%
_{ijk}+\hat{\sigma}_{i,j-1,k},\sigma _{ijk}+\sigma _{i,j-1,k}}\hat{\sigma}%
_{ijk}W_{i,j,k;i,j-1,k}\left( \sigma _{ijk}\sigma _{i,j-1,k}\rightarrow \hat{%
\sigma}_{ijk}\hat{\sigma}_{i,j-1,k}\right) \\
&=&\frac 1{2\left( b+K\right) }\left[ K\left( \sigma _{i,j,k+1}+\sigma
_{i,j+1,k}+\sigma _{i,j,k-1}+\sigma _{i-1,j,k}+\sigma _{i+1,j,k}+\sigma
_{ijk}+\sigma _{i,j-1,k}\right. \right. \\
&&\left. \left. -\sigma _{i,j-1,k+1}-\sigma _{i,j-2,k}-\sigma
_{i,j-1,k-1}-\sigma _{i+1,j-1,k}-\sigma _{i-1,j-1,k}\right) +b\left( \sigma
_{ijk}+\sigma _{i,j-1,k}\right) \right] ,
\end{eqnarray*}
\begin{eqnarray*}
&&\sum_{\hat{\sigma}_{ijk},\hat{\sigma}_{i,j,k+1}}\delta _{\hat{\sigma}%
_{ijk}+\hat{\sigma}_{i,j,k+1},\sigma _{ijk}+\sigma _{i,j,k+1}}\hat{\sigma}%
_{ijk}W_{i,j,k;i,j,k+1}\left( \sigma _{ijk}\sigma _{i,j,k+1}\rightarrow \hat{%
\sigma}_{ijk}\hat{\sigma}_{i,j,k+1}\right) \\
&=&\frac 1{2\left( b+K\right) }\left[ K\left( \sigma _{i+1,j,k}+\sigma
_{i,j,k-1}+\sigma _{i-1,j,k}+\sigma _{i,j-1,k}+\sigma _{i,j+1,k}+\sigma
_{ijk}+\sigma _{i,j,k+1}\right. \right. \\
&&\left. \left. -\sigma _{i+1,j,k+1}-\sigma _{i,j,k+2}-\sigma
_{i-1,j,k+1}-\sigma _{i,j+1,k+1}-\sigma _{i,j-1,k+1}\right) +b\left( \sigma
_{ijk}+\sigma _{i,j,k+1}\right) \right] ,
\end{eqnarray*}
\begin{eqnarray*}
&&\sum_{\hat{\sigma}_{ijk},\hat{\sigma}_{i,j,k-1}}\delta _{\hat{\sigma}%
_{ijk}+\hat{\sigma}_{i,j,k-1},\sigma _{ijk}+\sigma _{i,j,k-1}}\hat{\sigma}%
_{ijk}W_{i,j,k;i,j,k-1}\left( \sigma _{ijk}\sigma _{i,j,k-1}\rightarrow \hat{%
\sigma}_{ijk}\hat{\sigma}_{i,j,k-1}\right) \\
&=&\frac 1{2\left( b+K\right) }\left[ K\left( \sigma _{i+1,j,k}+\sigma
_{i,j,k+1}+\sigma _{i-1,j,k}+\sigma _{i,j-1,k}+\sigma _{i,j+1,k}+\sigma
_{ijk}+\sigma _{i,j,k-1}\right. \right. \\
&&\left. \left. -\sigma _{i+1,j,k-1}-\sigma _{i,j,k-2}-\sigma
_{i-1,j,k-1}-\sigma _{i,j+1,k-1}-\sigma _{i,j-1,k-1}\right) +b\left( \sigma
_{ijk}+\sigma _{i,j,k-1}\right) \right] .
\end{eqnarray*}
Substituting them into the time-evolving equation of the local magnetization
(\ref{Time-Q}), we get 
\begin{eqnarray}
\frac d{dt}q_{ijk}\left( t\right) &=&\frac 1{2\left( b+K\right) }b\left\{
\left[ \left( q_{i,j,k+1}-q_{ijk}\right) -\left( q_{ijk}-q_{i,j,k-1}\right)
\right] \right.  \nonumber \\
&&\left. +\left[ \left( q_{i+1,j,k}-q_{ijk}\right) -\left(
q_{ijk}-q_{i-1,j,k}\right) \right] +\left[ \left( q_{i,j+1,k}-q_{ijk}\right)
-\left( q_{ijk}-q_{i,j-1,k}\right) \right] \right\}  \nonumber \\
&&+\frac K{2\left( b+K\right) }\left[ 2\left(
2q_{i-1,j,k}-q_{i-1,j+1,k}-q_{i-1,j-1,k}\right) +\left(
2q_{i-1,j,k}-q_{ijk}-q_{i-2,j,k}\right) \right.  \nonumber \\
&&+2\left( 2q_{i+1,j,k}-q_{i+1,j+1,k}-q_{i+1,j-1,k}\right) +\left(
2q_{i+1,j,k}-q_{ijk}-q_{i+2,j,k}\right)  \nonumber \\
&&+2\left( 2q_{i,j-1,k}-q_{i,j-1,k+1}-q_{i,j-1,k-1}\right) +\left(
2q_{i,j-1,k}-q_{ijk}-q_{i,j-2,k}\right)  \nonumber \\
&&+2\left( 2q_{i,j+1,k}-q_{i,j+1,k+1}-q_{i,j+1,k-1}\right) +\left(
2q_{i,j+1,k}-q_{ijk}-q_{i,j+2,k}\right)  \nonumber \\
&&+2\left( 2q_{i,j,k-1}-q_{i-1,j,k-1}-q_{i+1,j,k-1}\right) +\left(
2q_{i,j,k-1}-q_{ijk}-q_{i,j,k-2}\right)  \nonumber \\
&&\left. +2\left( 2q_{i,j,k+1}-q_{i+1,j,k+1}-q_{i-1,j,k+1}\right) +\left(
2q_{i,j,k+1}-q_{ijk}-q_{i,j,k+2}\right) \right] .  \label{3d-Q-result}
\end{eqnarray}
With lattice constant $a$ we can transform the above equation to be 
\begin{eqnarray}
\frac d{dt}q\left( t\right) &=&\frac{a^2}{2\left( b+K\right) }b\left( \nabla
_x^2+\nabla _y^2+\nabla _z^2\right) q\left( t\right)  \nonumber \\
&&-\frac{a^2}{2\left( b+K\right) }K\left[ \left( 2\nabla _y^2+\nabla
_x^2\right) +\left( 2\nabla _y^2+\nabla _x^2\right) +\left( 2\nabla
_z^2+\nabla _y^2\right) \right.  \nonumber \\
&&\left. +\left( 2\nabla _z^2+\nabla _y^2\right) +\left( 2\nabla _x^2+\nabla
_z^2\right) +\left( 2\nabla _x^2+\nabla _z^2\right) \right] q\left( t\right)
\nonumber \\
&=&\frac{3a^2}{b+K}\left( \frac b6-K\right) \nabla ^2q\left( t\right) .
\label{3d-diffusion}
\end{eqnarray}
It is of the form of a diffusion equation 
\[
\frac{dq\left( t\right) }{dt}=D\nabla ^2q\left( t\right) , 
\]
where 
\begin{equation}
D=\frac{3a^2}{b+K}\left( \frac b6-K\right) a^2.
\end{equation}

With the same treatment, we can easily obtain the diffusion equation of one
and two dimensional kinetic Gaussian model. Here we give only the results, 
\begin{equation}
\frac d{dt}q_\alpha \left( t\right) =\frac{da^2}{b+K}\left( \frac b{2d}%
-K\right) \nabla ^2q,
\end{equation}
where $d$ is the system dimensionality. As we know that the critical point
for $d$-dimensional Gaussian model is $K_c=J/k_BT_c=b/2d$, thus, the
diffusion equation we obtained reveals that the diffusion will get much
slower when it is near the critical point. The linear equations we obtained
for a single spin can be directly solved, but the solution of the diffusion
equations may already give us satisfying information. For example, in $1D$
case, it is 
\[
q\left( x,t\right) =\frac 1{2\sqrt{D\pi t}}\int_{-\infty }^\infty q\left(
\xi ,0\right) e^{-\frac{\left( x-\xi \right) ^2}{4Dt}}d\xi . 
\]
In a specific example of the diffusion of a Gaussian type packet, $q\left(
\xi ,0\right) =e^{-\xi ^2}$, one will obtain 
\begin{equation}
q\left( x,t\right) =\sqrt{\frac 1{1+t/\tau }}\exp \left[ -\frac{x^2}{%
1+t/\tau }\right] ,
\end{equation}
where 
\begin{equation}
\tau =\frac 1{4D}=\left[ \frac{4a^2}{b+K}\left( \frac b2-K\right) \right]
^{-1}
\end{equation}
is the relaxation time. When $K\rightarrow K_c=b/2$, $D\rightarrow 0$ and $%
\tau \rightarrow \infty $, and this is a typical critical slowing down
phenomenon. With the correlation length critical exponent $\nu =1/2$ and the
following dynamical scaling hypotheses 
\[
\xi \sim \left| T-T_c\right| ^{-\nu },\tau \sim \xi ^z, 
\]
one can obtain the dynamic critical exponent $z=2$. The same result can be
obtained for $2D$ and $3D$ model. In earlier study\cite{Zhu-1} the same
result $z=1/v=2$ for any dimensionality was obtained with Glauber-type
mechanism. Thus we find that, in Kinetic Gaussian model, {\it the critical
dynamic exponent is independent of space dimensionality and the dynamic
mechanism}.

Before conclusion, we may note that actually, although a systematic
formulation has been lacked, efforts towards wider application of Kawasaki's
dynamics have never been stopped. For example, in a Monte Carlo simulation
of the three-dimensional ferromagnetic Heisenberg model\cite{zp-Z}, Zhang
has suggested that two neighboring spins, $\sigma _i$ and $\sigma _j$, may
rotate with their conserved sum being the axis. He found that the new scheme
enabled the system to evolve to thermodynamic equilibrium faster, and
commented that it might be more favorable in reality. This is just an
successful exploration of the spin-pair redistribution mechanism, and there
are many other such examples in non-equilibrium statistics, though assuming
different forms.

To summarize, in this article, we presented a systematic formulation of the
Kawasaki-type dynamics: spin-pair redistribution. As a natural
generalization of Kawasaki's exchange mechanism, the new dynamics gives the
system more freedom while keeping the order-parameter conserved. We have
analyzed the redistribution in physics, given the master equation and a
normalized redistribution probability determined by the {\it heat Boltzmann
factor. }The presentation of this probability, being the key of the whole
formulation, makes the mechanism mathematically well-organized and
physically meaningful. As mentioned above, in many works focused on specific
lattice systems there are already ideas of ''redistribution'', and these
efforts often turned to be rather fruitful. However without a universal
theoretical foundation, these works still remained tentative to some degree.
This article compactly and systematically provides this foundation, upon
which the generalized Kawasaki dynamics becomes universal and can be {\it %
directly} applied to microscopic systems. Without any extra requirements, it
has its advantage compared with some earlier approaches covering the same
ground, such as numerical Ginzberg Landau approaches.

The formulation is compact in mathematics, while on the other hand it is
also quite open. People are able to introduce other elements into it. For
example, for decades there have been attempts of the application of a
combined mechanism, Glauber type and Kawasaki type both with a probability%
\cite{GK}. We can easily give the mathematical form of the combined
mechanism, single-spin transition plus spin-pair redistribution. In fact for
Gaussian model we have already obtained exact and physically clear results,
which will be reported in later paper as further application. For another
example we may note the recent interest in the small-world network\cite{SWN}
effect on transportation. One can also modify the redistribution mechanism
and directly study these effects in non-equilibrium statistical dynamics.

The formulation of the mechanism is the chief purpose of this article. As an
example of its applications we obtained exact diffusion equations in kinetic
Gaussian model, and a temperature-dependent diffusion coefficient which
becomes zero at the critical point. We observed critical slowing down
phenomenon in diffusion process, and found that at least in this specific
case the critical dynamic exponent is independent of space dimensionality
and the dynamic mechanism.

Up to this point we have successfully generalized Glauber's single-spin
flipping mechanism to single-spin transition mechanism, and Kawasaki's
spin-pair exchange mechanism to spin-pair redistribution mechanism. These
two generalizations are of similar mathematical form and become counterparts
of each other in nonconserved dynamics and conserved dynamics respectively.

The diffusion in Gaussian model studied in this article is only an example
of the problems previously we were unable to treat. Based on the
redistribution mechanism (or transition), one can also easily write the
master equation and redistribution probability for Potts model, XY model,
Heisenberg model and many other types (in principle arbitrary), and obtain
the evolution of local magnetization (and correlation function, etc.) in
these systems. Though exact treatment may be difficult, other methods can be
later applied based on this foundation. At the same time this mechanism can
be almost directly applied to conserved processes other than those in
spin-lattice models. For example, it can be used to study the relaxation of
granular material under shaking (mass conserved), or the activities of
particles in space (particle number conserved). A feasible way may be like
this: first we divide the system into small units and define the parameter
(and its range) and the system Hamiltonian. The redistribution is not
necessarily limited to neighboring units (we can modify the master equation
accordingly) and the probability may assume a form other than the heat
Boltzmann factor (but it has to be normalized). Then we write the master
equation and the evolving equations of the parameters we are interested in.
This serves to unite the various conserved processes in microscopic, place
them under the control of a universal mechanism, and provide the basis for
further treatment, either exact, approximation, or Monte Carlo. The same is
for the single-spin transition mechanism. We think it is very important for
the general progress in non-equilibrium statistical dynamics.

\acknowledgments 

This work was supported by the National Natural Science Foundation of China
under Grant No. 10075025.

\appendix 

\section{Equation derivation}

\label{App-A}

\begin{eqnarray*}
\frac d{dt}q_k\left( t\right) &=&\frac d{dt}\sum_{\left\{ \sigma \right\}
}\sigma _kP\left( \left\{ \sigma \right\} ;t\right) \\
&=&\sum_{\left\{ \sigma \right\} }\sum_{\left\langle j,l\right\rangle }\sum_{%
\hat{\sigma}_j,\hat{\sigma}_l}\left[ -\sigma _kW_{jl}\left( \sigma _j\sigma
_l\rightarrow \hat{\sigma}_j\hat{\sigma}_l\right) P\left( \left\{ \sigma
\right\} ;t\right) \right. \\
&&\left. +\sigma _kW_{jl}\left( \hat{\sigma}_j\hat{\sigma}_l\rightarrow
\sigma _j\sigma _l\right) P\left( \left\{ \sigma _{i\neq j,l}\right\} ,\hat{%
\sigma}_j,\hat{\sigma}_l;t\right) \right]
\end{eqnarray*}
Taking summation over all $\left\{ \sigma \right\} $, one will find the
terms of those pairs unrelated to $\sigma _k$ cancel with each other. So one
will only have 
\begin{eqnarray*}
&&\frac d{dt}q_k\left( t\right) \\
&=&\sum_{\left\{ \sigma \right\} }\sum_w\sum_{\hat{\sigma}_k,\hat{\sigma}%
_{k+w}}\left[ -\sigma _k\left( W_{k,k+w}\left( \sigma _k\sigma
_{k+w}\rightarrow \hat{\sigma}_k\hat{\sigma}_{k+w}\right) \right) P\left(
\left\{ \sigma \right\} ;t\right) \right. \\
&&\left. +\sigma _kW_{k,k+w}\left( \hat{\sigma}_k\hat{\sigma}%
_{k+w}\rightarrow \sigma _k\sigma _{k+w}\right) P\left( \sigma _1,\ldots ,%
\hat{\sigma}_k,\hat{\sigma}_{k+w},\ldots ,\sigma _N;t\right) \right] \\
&=&-2dq_k\left( t\right) +\sum_{\left\{ \sigma \right\} }\sum_w\left[ \sum_{%
\hat{\sigma}_k,\hat{\sigma}_{k+w}}\sigma _kW_{k,k+w}\left( \hat{\sigma}_k%
\hat{\sigma}_{k+w}\rightarrow \sigma _k\sigma _{k+w}\right) P\left( \sigma
_1,\ldots ,\hat{\sigma}_k,\hat{\sigma}_{k+w},\ldots ,\sigma _N;t\right)
\right] \\
&=&-2dq_k\left( t\right) \\
&&+\sum_w\sum_{\sigma _1,\ldots ,\sigma _k,\sigma _{k+w},\hat{\sigma}_k,\hat{%
\sigma}_{k+w},\ldots ,\sigma _N}\sigma _kW_{k,k+w}\left( \hat{\sigma}_k\hat{%
\sigma}_{k+w}\rightarrow \sigma _k\sigma _{k+w}\right) P\left( \sigma
_1,\ldots ,\hat{\sigma}_k,\hat{\sigma}_{k+w},\ldots ,\sigma _N;t\right) \\
&=&2dq_k\left( t\right) +\sum_{\left\{ \sigma \right\} }\sum_w\left[ \sum_{%
\hat{\sigma}_k,\hat{\sigma}_{k+w}}\hat{\sigma}_kW_{k,k+w}\left( \sigma
_k\sigma _{k+w}\rightarrow \hat{\sigma}_k\hat{\sigma}_{k+w}\right) \right]
P\left( \left\{ \sigma \right\} ;t\right) .
\end{eqnarray*}

\section{Calculational details}

\label{App-B}

We give the details for one of them, and the remaining terms follow the same
way. The spin pair redistribution probability can be expressed as 
\begin{eqnarray*}
&&W_{i,j,k;i+1,j,k}\left( \sigma _{ijk}\sigma _{i+1,j,k}\rightarrow \hat{%
\sigma}_{ijk}\hat{\sigma}_{i+1,j,k}\right) \\
&=&\frac 1{Q_{i,j,k;i+1,j,k}}\delta _{\hat{\sigma}_{ijk}+\hat{\sigma}%
_{i+1,j,k},\sigma _{ijk}+\sigma _{i+1,j,k}} \\
&&\times \exp \left\{ K\left[ \hat{\sigma}_{ijk}\left( \sigma
_{i-1,j,k}+\sigma _{i,j+1,k}+\sigma _{i,j-1,k}+\sigma _{i,j,k-1}+\sigma
_{i,j,k+1}\right) +\hat{\sigma}_{ijk}\hat{\sigma}_{i+1,j,k}\right. \right. \\
&&\left. \left. +\hat{\sigma}_{i+1,j,k}\left( \sigma _{i+2,j,k}+\sigma
_{i+1,j+1,k}+\sigma _{i+1,j-1,k}+\sigma _{i+1,j,k+1}+\sigma
_{i+1,j,k-1}\right) \right] \right\} ,
\end{eqnarray*}
and there are similar expressions for $W_{i,j,k;i-1,j,k}$, $W_{i,j,k;i,j\pm
1,k}$, and $W_{i,j,k;i,j,k\pm 1}$. Because the spins take continuous value,
the summation for spin value turns into the integration 
\[
\sum_\sigma \rightarrow \int_{-\infty }^\infty f\left( \sigma \right)
d\sigma ; 
\]
and then the normalization factor 
\begin{eqnarray*}
&&Q_{i,j,k;i+1,j,k} \\
&=&\frac b{2\pi }\int\limits_{-\infty }^{+\infty }\int\limits_{-\infty
}^{+\infty }d\hat{\sigma}_{ijk}d\hat{\sigma}_{i+1,j,k}\delta _{\hat{\sigma}%
_{ijk}+\hat{\sigma}_{i+1,j,k},\sigma _{ijk}+\sigma _{i+1,j,k}} \\
&&\times \exp \left\{ K\left[ \hat{\sigma}_{ijk}\left( \sigma
_{i,j+1,k}+\sigma _{i-1,j,k}+\sigma _{i,j-1,k}+\sigma _{i,j,k-1}+\sigma
_{i,j,k+1}\right) +\hat{\sigma}_{ijk}\hat{\sigma}_{i+1,j,k}\right. \right. \\
&&\left. \left. +\hat{\sigma}_{i+1,j,k}\left( \sigma _{i+1,j+1,k}+\sigma
_{i+2,j,k}+\sigma _{i+1,j-1,k}+\sigma _{i+1,j,k+1}+\sigma
_{i+1,j,k-1}\right) \right] -\frac b2\hat{\sigma}_{ijk}^2-\frac b2\hat{\sigma%
}_{i+1,j,k}^2\right\} \\
&=&\frac b{2\pi }\sqrt{\frac \pi {b+K}}\exp \left[ K\left( \sigma
_{ijk}+\sigma _{i+1,j,k}\right) \right. \\
&&\left. \times \left( \sigma _{i+1,j+1,k}+\sigma _{i+2,j,k}+\sigma
_{i+1,j-1,k}+\sigma _{i+1,j,k+1}+\sigma _{i+1,j,k-1}\right) -\frac b2\left(
\sigma _{ijk}+\sigma _{i+1,j,k}\right) ^2\right] \\
&&\times \exp \left\{ \frac 1{4\left( b+K\right) }\left[ K\left( \sigma
_{i,j+1,k}+\sigma _{i-1,j,k}+\sigma _{i,j-1,k}+\sigma _{i,j,k-1}+\sigma
_{i,j,k+1}+\sigma _{ijk}+\sigma _{i+1,j,k}\right. \right. \right. \\
&&\left. \left. \left. -\sigma _{i+1,j+1,k}-\sigma _{i+2,j,k}-\sigma
_{i+1,j-1,k}-\sigma _{i+1,j,k+1}-\sigma _{i+1,j,k-1}\right) +b\left( \sigma
_{ijk}+\sigma _{i+1,j,k}\right) ^2\right] \right\} .
\end{eqnarray*}
Thus this combined term in Eqs.(\ref{Time-Q}) becomes 
\begin{eqnarray*}
&&\sum_{\hat{\sigma}_{ijk},\hat{\sigma}_{i+1,j,k}}\hat{\sigma}%
_{ijk}W_{i,j,k;i+1,j,k}\left( \sigma _{ijk}\sigma _{i+1,j,k}\rightarrow \hat{%
\sigma}_{ijk}\hat{\sigma}_{i+1,j,k}\right) \\
&=&\frac b{2\pi }\frac 1{Q_{i,j,k;i+1,j,k}}\int\limits_{-\infty }^{+\infty
}\int\limits_{-\infty }^{+\infty }d\hat{\sigma}_{ijk}d\hat{\sigma}%
_{i+1,j,k}\delta _{\hat{\sigma}_{ijk}+\hat{\sigma}_{i+1,j,k},\sigma
_{ijk}+\sigma _{i+1,j,k}} \\
&&\times \hat{\sigma}_{ijk}\exp \left\{ K\left[ \hat{\sigma}_{ijk}\left(
\sigma _{i,j+1,k}+\sigma _{i-1,j,k}+\sigma _{i,j-1,k}+\sigma
_{i,j,k-1}+\sigma _{i,j,k+1}\right) +\hat{\sigma}_{ijk}\hat{\sigma}%
_{i+1,j,k}\right. \right. \\
&&\left. \left. +\hat{\sigma}_{i+1,j,k}\left( \sigma _{i+1,j+1,k}+\sigma
_{i+2,j,k}+\sigma _{i+1,j-1,k}+\sigma _{i+1,j,k+1}+\sigma
_{i+1,j,k-1}\right) \right] -\frac b2\hat{\sigma}_{ij}^2-\frac b2\hat{\sigma}%
_{i+1,j}^2\right\} \\
&=&\frac 1{2\left( b+K\right) }\left[ K\left( \sigma _{i,j+1,k}+\sigma
_{i-1,j,k}+\sigma _{i,j-1,k}+\sigma _{i,j,k-1}+\sigma _{i,j,k+1}+\sigma
_{ijk}+\sigma _{i+1,j,k}\right. \right. \\
&&\left. \left. -\sigma _{i+1,j+1,k}-\sigma _{i+2,j,k}-\sigma
_{i+1,j-1,k}-\sigma _{i+1,j,k+1}-\sigma _{i+1,j,k-1}\right) +b\left( \sigma
_{ijk}+\sigma _{i+1,j,k}\right) \right] .
\end{eqnarray*}


\begin{references}
\bibitem{Glauber}  R. J. Glauber, J. Math. Phys. {\bf 4}, 294 (1963).

\bibitem{Kawasaki}  K. Kawasaki, Phys. Rev. {\bf 145}, 224 (1966).

\bibitem{Ising-1}  K. E. Bassler and Z. Racz, Phys. Rev. Lett. {\bf 73},
1320 (1994)

\bibitem{Ising-2}  P. Fratzl and O. Penrose, Acta Materialia, {\bf 43}, 2921
(1995); {\bf 44}, 3227 (1996)

\bibitem{Lat-Gas-1}  Y. He and R. B. Pandey, Phys. Rev. Lett. {\bf 71}, 565
(1993).

\bibitem{Lat-Gas-2}  A. Szolnoki, G. Szabo and O. G. Mouritsen, Phys. Rev. E 
{\bf 55, }2255 (1997).

\bibitem{Lat-Gas-3}  S. Weinketz, Phys. Rev. E {\bf 58}, 159 (1998).

\bibitem{BEG-1}  M. Porta, C. Frontera, E. Vives and T. Castan, Phys. Rev. B 
{\bf 56}, 5261 (1997).

\bibitem{BEG-2}  J. F. F. Mendes., S.Cornell, M. Droz and E. J. S. Lage, J.
Phys. A {\bf 25}, 73 (1992).

\bibitem{Others-1}  R. A. Denny and M. V. Sangaranarayanan, Chem. Phys.
Lett. {\bf 239}, 131 (1995).

\bibitem{Others-2}  P. Fratzl and O. Penrose, Phys. Rev. B {\bf 55} R6101
(1997).

\bibitem{Zhu-1}  Jian-Yang Zhu and Z. R. Yang, Phys. Rev. E {\bf 59, }1551
(1999).

\bibitem{Zhu-2}  Jian-Yang Zhu and Z. R. Yang, Phys. Rev. E {\bf 61, }210
(2000), Phys. Rev. E {\bf 61, }6219 (2000).

\bibitem{zp-Z}  Zhengping Zhang, Phys. Rev. E {\bf 51, }4155 (1995).

\bibitem{GK}  There has been continuing study in this direction. Some of the
recent works on Ising model are: Ma Yuqiang and Liu Jiwen, Phys. Lett. A 
{\bf 238}, 159 (1998); Yu-Qiang Ma, Ji-Wen Liu and W. Figueiredo, Phys. Rev.
E {\bf 57, }3625 (1998); Attila Szolnoki, Phys. Rev. E {\bf 62}, 7466
(2000); B. C. S. Grandi and W. Figueiredo, Phys. Rev. E {\bf 53}, 5484
(1996), {\bf 54}, 4722 (1996), {\bf 56}, 5240 (1997).

\bibitem{SWN}  It refers to the introduction of a certain amount of random
long-range connections into an initially regular network. There are a lot of
interesting and unexpected characteristics in such structures. D. J. Watts
and S. H. Strogatz, Nature {\bf 393}, 440 (1998).
\end{references}
\end{document}